\documentstyle[twoside,fleqn,jhuwkshp,psfig]{article}
\newcommand{\etal}{{\it et al.}}

\newcommand{\ginga}{{\it Ginga}}

\begin{document}

\title{REFLECTION AT LARGE DISTANCE\\ FROM THE CENTRAL ENGINE IN SEYFERTS}
 
\author{J. Malzac \address{Osservatorio Astronomico di Brera, via Brera
28, 20121, Milano,Italy}, P.O. Petrucci $^{a}$ \address{Laboratoire
d'Astrophysique, Observatoire de Grenoble, BP 53X, 38041 Grenoble Cedex,
France} }

\begin{abstract}
We consider the possibility that most of the reflection component,
 observed in the hard X-ray spectra of Seyfert galaxies, could be formed
 on an extended medium, at large distance from the central source of
 primary radiation (e.g. on a torus). Then, the reflector cannot respond
 to the rapid fluctuations of the primary source. The observed reflected
 flux is controlled by the time-averaged primary spectrum rather than the
 instantaneous (observed) one. We show that this effect strongly
 influence the spectral fits parameters derived under the assumption of a
 reflection component consistent with the primary radiation.  We find
 that a pivoting primary power-law spectrum with a nearly constant
 Comptonised luminosity may account for the reported correlation between
 the reflection amplitude $R$ and the spectral index $\Gamma$.
\end{abstract}

\maketitle

\section{Introduction}

In radio-quiet active galactic nuclei, large amounts of cold material are
thought to reside at large distance from the central engine.  In
particular, the unified scheme (Antonucci 1993) postulates a large scale
torus (with radius larger than $10^{16}$ cm).  Such {distant cold
material (hereafter DCM)} may imprint strong reflection features in the
hard X-ray spectrum of Seyfert 1 galaxies (Ghisellini, Haardt \& Matt
1994; Krolik, Madau \& \.Zycki 1994).  It has been argued that the
reflection produced by an irradiated torus would be sufficient to account
for the typical observed reflection spectra, without need for reflecting
material in the inner parts of the accretion flow (e.g. disc reflection).

In Compton thick Seyfert 2s, the observed reflection dominated spectra
are generally interpreted by DCM reflection with the primary emission
obscured presumably by the same outer material (e.g. Reynold \etal, 1994;
Matt \etal, 2000).

In Seyfert 1s, the detection of a narrow iron line component (Lubinski \&
Zdziarski 2000) is also suggestive of reflection on DCM.  Probably, the
best evidence for significant reflection on DCM in a Seyfert 1, comes
from the BeppoSax and RXTE observation of NGC4051 in a very low flux
state. Both spectral and timing data were found consistent with the
interpretation that the source had switched off, leaving only a spectrum
of pure reflection from DCM at distances larger than $10^{17}$ cm
(Guainazzi \etal 1998, Uttley \etal 1999).

The aim of this paper is to emphasize some observable effects introduced
 by DCM reflection and discuss these effects in the context of the
 reported $R$--$\Gamma$ correlation (Zdziarski, Lubi\~{n}ski \& Smith
 1999).

\section{The measured reflection amplitude $R$}

When analyzing the X-ray data, the possible contribution from DCM
reflection is not considered a priori.  The data are generally
interpreted in the framework of reflection in the vicinity of the hard
X-ray source, possibly on the accretion disc.  Usually, in spectral fits,
the shape of the reflection component is computed assuming that the
\emph{observed} primary spectrum illuminates an infinite disc (e.g. {\sc
pexrav} model in {\sc xpec}, Magdziarz \& Zdziarski 1995).  The
normalization of the reflected spectrum is then tuned in order to fit the
observed spectrum.  The results of this fitting procedure provides the
reflection amplitude $R$.  $R$ is normalized so that $R=1$ in the case of
an isotropic source above an infinite reflecting plane. $R$ is often
considered as an estimate of $\Omega/2\pi$ where $\Omega$ is the solid
angle subtended by the reflector as seen from the isotropic X-ray source.

\section{Effects of a distant reflector}

Obviously, any contribution from a remote structure leads to an increase
of $R$ and may lead to an overestimate of the disc reflection.  This may
explain the very large $R$ coefficients $\sim 2$, measured in some
Seyfert 1s, which are difficult to reconcile with disc reflection (see
however Beloborodov 1999; Malzac, Beloborodov \& Poutanen 1999).  In
addition, the nuclei of Seyfert galaxies are known to harbor a
significant variability on very short time scale ( $<$ 1 day, see
e.g. Nicastro \etal 2000; Nandra \etal 2000; Chiang \etal 2000).  Due to
its extended structure the remote reflector cannot respond to the rapid
fluctuations of the primary X-ray flux.  \emph{The reflected component
from the DCM is thus likely to correspond to the time-averaged incident
flux rather than to the instantaneous (i.e. observed) one.}  Thus flux
changes may induce a significant variation in the $R$ value derived from
the spectral fits.  A flux lower than the average enhances the apparent
reflection, on the other hand, a larger flux may reduce $R$ down to zero.
Then values of $R$ as low or large as required by the data can be easily
produced. This kind of effect, when important, makes the geometrical
interpretation of $R$ extremely misleading.  \emph{Trying to disentangle
the temporal and geometrical effects is extremely difficult} and requires
very long observations with time resolved spectral analysis.

\section{Is distant reflection 
consistent with a $R$-$\Gamma$ correlation ?}

A significant contribution from a remote reflector seems, at first sight,
in contradiction with the reported correlation between $R$ and the
spectral slope $\Gamma$ (Zdziarski, Lubi\~{n}ski \& Smith, 1999).  This
correlation is observed in sample of sources as well as in the time
evolution of individual sources.  The measured $R$ tends to be larger in
softer sources.  The usual interpretation of the correlation invoke the
feedback from reprocessed radiation emitted by the reflector itself. It
thus absolutely requires a close reflector.  In the context of DCM
reflection, it is difficult to understand why the reflection contribution
from DCM should be more important in objects with softer spectra.

It has been however suggested by Nandra \etal (2000) that, in the case of
reflection on DCM, \emph{ rapid spectral changes of the primary emission
can strongly affect the measured $R$.  The $R$ vs $\Gamma$ relation then
depends on the specific spectral variability mode of the sources.}  It
may possibly produce a correlation between $R$ and $\Gamma$.

In several sources such as NGC5548 (Nicastro \etal 2000, Petrucci \etal
2000), the short time-scale variability is consistent with fluctuations
of the X-ray spectral slope with a nearly constant comptonised
luminosity, i.e. the spectrum is mainly pivoting.  This can happen for
example if the UV luminosity entering the hot Comptonising plasma changes
with a constant heating rate in the hot plasma (e.g. Malzac \& Jourdain
2000).  In the following we investigate the $R$ vs $\Gamma$ dependence
for this variability mode.

\section{Modeling method}

We assume that the time average primary spectrum seen by the DCM can be
represented by a power law with a photon index $\Gamma=1.9$ and an
exponential cut-off at 200 keV. Neglecting any disc reflection, we
further assume that the DCM geometry is such that the time averaged
primary spectrum yield a reflection coefficient $R=0.5$.  As a first
order approximation, the shape of the reflection spectrum is computed
using the {\sc pexrav} procedure, i.e. assuming a slab reflector.  Fixing
the cut-off energy at 200 keV, we produced a set of instantaneous primary
spectra for several photon indices $\Gamma$ spanning the observed range
1.4--2.2. The 1 keV normalization was tuned so that the 0.01--100 keV
luminosity was identical for all spectra.  We then added the reflection
component, produced as described above, to these instantaneous spectra,
and fitted the resulting spectra with {\sc pexrav} in the 2-30 keV
range. We also fitted the simulated spectra in the 2--100 keV range. In
both energy ranges, the derived best fit parameters were very similar.
Similarly, we estimated the effects of variability on the measured
equivalent width (EW) of the iron line.  We modeled the intrinsic iron
line by a Gaussian with an intrinsic width $\sigma=0.1$ eV and a total
flux normalized so that the EW is 50 eV for the $\Gamma=1.9$ time
averaged primary spectra (i.e. roughly consistent with a reflection
amplitude $R=0.5$; George \& Fabian 1991).

\section{Results}

\begin{figure}[t] 
\vspace{10pt}
\centerline{\psfig{file=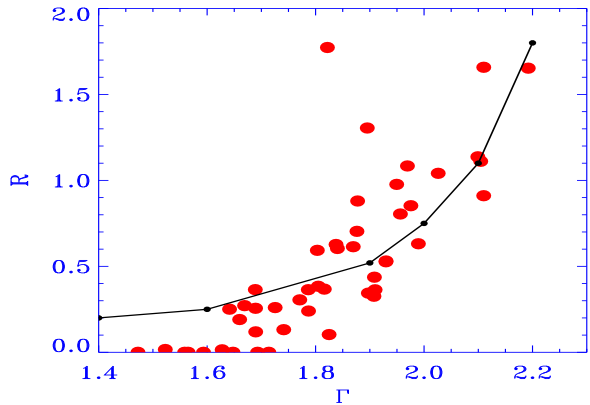,width=3.0in,height=2.0in}}
\caption{The $R$--$\Gamma$ correlation obtained assuming a primary
powerlaw spectrum with a varying photon index, at constant 0.01-100 keV
luminosity and a fixed reflected flux (solid line; see text).  The
simulated spectra were fitted using {\sc pexrav} in the 2-30 keV range.
The circles show the \ginga data of Zdziarski \etal 1999.  The errors are
omitted.  }\label{fig:smallfig1}
\end{figure}                                                                             
\begin{figure}[t] 
\vspace{10pt}
\centerline{\psfig{file=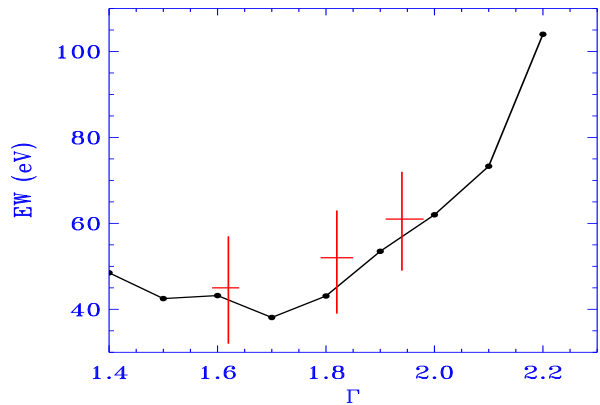,width=3.0in,height=2.0in}}
\caption{ The iron line equivalent width as a function of $\Gamma$
assuming a primary Comptonised spectrum with a varying photon index a
constant 0.01-100 keV luminosity and a fixed line flux (solid line; see
text). The crosses show the data for the three averaged spectra of
Lubi\~{n}ski \& Zdziarski (2000) }\label{fig:smallfig2}
\end{figure}    
\subsection{The $R$--$\Gamma$ correlation}

The $R$-$\Gamma$ relation resulting from our modeling procedure is shown
in Fig.~1. The assumed variability mode (i.e. variability at constant
flux) results in a spectrum pivoting at energies lower than 10 keV i.e.
below the energy range were most of reflection is produced.  Thus, when
the spectrum is hard the observed primary flux in the 10--30 keV band is
larger, and $R$ is reduced, on the other hand when the spectrum is soft
the primary flux in this energy range is enhanced and $R$ is lower. It
thus produces \emph{a positive correlation between $R$ and $\Gamma$}.
Fig.~1 shows that \emph{ the produced correlation qualitatively match the
observed $R$-$\Gamma$ correlation observed in the sample of \ginga data
from Zdziarski \etal 1999}.  Note that a spectral pivot above 10-20 keV
would produce the an anti-correlation.

\subsection{The iron line equivalent width vs $\Gamma$ relation}

Fluctuations of the primary continuum do affect the measured EW.  The
resulting dependence of the measured equivalent width with $\Gamma$ is
plotted on Fig~2.  The EW represents the amount of reflection at 6.4 keV
i.e. at a lower energy than the reflection bump.  The correlation
obtained for EW is weaker than for $R$ (actually we even get an
anti-correlation at low $\Gamma$).

Using a large ASCA sample, Lubi\'nski \& Zdziardski (2000) produced 3
average spectra of Seyfert 1s grouped according to increasing spectral
index. For each spectra they could detect a narrow Fe line component that
they attribute to reflection on a torus.  \emph{Their estimates for the
narrow line EW appear to be consistent with the results from our
modeling} as show in Fig~2.

\section{Conclusion}

We showed that the effect of the remote cold material may in principle
account simultaneously for the observed correlation between the $R$ and
$\Gamma$, and the observed narrow line component which is almost
independent of $\Gamma$.  It should be stressed however, that the
comparisons of our model with the $R$-$\Gamma$ correlation observed in a
{\emph sample} of sources, implicitly assumes that all the sources
present a similar variability mode (i.e. variability at constant flux),
time average properties as well as geometry of the DCM.  This is probably
oversimplifying (as suggested by the spread of the data points in the
$R$-$\Gamma$ plane).  In particular in several sources a correlation
between the 2--10 keV flux and spectral slope is reported (e.g. Chiang
\etal 2000), while our variability assumptions lead to an
anti-correlation.  Concerning the $R$-$\Gamma$ correlation we implicitly
assumed that the reflection on inner cold material (e.g. disc reflection)
is negligible in all sources. This is probably not the case since the
observed iron lines also present a broad component which is unlikely to
be produced far away from the central engine.

Given these uncertainties, we conclude that beside the extreme DCM
dominated case considered here for illustration, \emph{reflection arising
from the DCM could be important in many sources}, making any attempt to
constrain the physical properties and the geometry of the central engine
with reflection measurements very difficult.


\normalsize

\section*{ACKNOWLEDGEMENTS}
This work  was partially supported by the Italian MURST through grant COFIN98-02-15-41 (J.M.) and by the European Commission under contract number ERBFMRX-CT98-0195 (TMR network "Accretion onto black holes, compact stars and
protostars").

\end{document}